# The Infrared Imaging Spectrograph (IRIS) for TMT: Instrument Overview

James E. Larkin[*a], Anna M. Moore[b], Elizabeth J. Barton[c], Brian Bauman[d], Khanh Bui[b], John Canfield[a], David Crampton[e], Alex Delacroix[b], Murray Fletcher[e], David Hale[b], David Loop[e], Cyndie Niehaus[a], Andrew C. Phillips[f], Vladimir Reshetov[e], Luc Simard[e], Roger Smith[b], Ryuji Suzuki[g,h], Tomonori Usuda[g], and Shelley A. Wright[c,i]

[a]Department of Physics and Astronomy, University of California, Los Angeles, CA 90095-1547;
[b]Caltech Optical Observatories, 1200 E California Blvd., Mail Code 11-17, Pasadena, CA 91125;
[c]Department of Physics & Astronomy, University of California, Irvine, 2158 Frederick Reines Hall, Irvine, CA 92697-4575;
[d]Lawrence Livermore National Laboratory, 7000 East Ave., M/S L-210, Livermore, CA 94550;
[e]Herzberg Institute of Astrophysics (HIA), National Research Council Canada, 5071 W Saanich Rd, Victoria, V9E 2E7;
[f]University of California Observatories, CfAO, University of California, 1156 High St., Santa Cruz, CA, 95064;
[g]Subaru Telescope, National Astronomical Observatory of Japan, 650 North A'ohoku Place, Hilo, HI 96720;
[h]Thirty Meter Telescope Observatory Corporation, 2632 E. Washington Blvd., Pasadena, CA, 91107
[i]University of California, Berkeley, Astronomy Department, 601 Campbell Hall, Berkeley, CA 94720

## ABSTRACT

We present an overview of the design of IRIS, an infrared (0.85 - 2.5 micron) integral field spectrograph and imaging camera for the Thirty Meter Telescope (TMT). With extremely low wavefront error (<30 nm) and on-board wavefront sensors, IRIS will take advantage of the high angular resolution of the narrow field infrared adaptive optics system (NFIRAOS) to dissect the sky at the diffraction limit of the 30-meter aperture. With a primary spectral resolution of 4000 and spatial sampling starting at 4 milliarcseconds, the instrument will create an unparalleled ability to explore high redshift galaxies, the Galactic center, star forming regions and virtually any astrophysical object. This paper summarizes the entire design and basic capabilities. Among the design innovations is the combination of lenslet and slicer integral field units, new 4Kx4k detectors, extremely precise atmospheric dispersion correction, infrared wavefront sensors, and a very large vacuum cryogenic system.

## 1. INTRODUCTION

IRIS (InfraRed Imaging Spectrograph) is a first light instrument currently being developed for the Thirty Meter Telescope (TMT)[1]. It works with the advanced adaptive optics system NFIRAOS[2] and has integrated on-instrument wavefront sensors (OIWFS)[3]. IRIS combines a powerful "wide field" imager[4] and an integral field spectrograph[5] both covering wavelengths from 0.84 μm to 2.4 μm. Both science instruments are built around expected 4K by 4K HgCdTe detectors from Teledyne (Hawaii 4RG)[6]. The imager has 4 milliarcsecond (mas) pixels and a total field of view of 16.4 arcseconds on a side. The spectrograph has four plate scales ranging from 4 to 50 mas and at least 3900 spatial sampling points (spaxels) in its primary modes. A spectral resolution of 4000 is provided for all broad bandpasses and higher spectral resolutions are supported for select narrow band regions. It will be capable of working spatially at the diffraction limit of the telescope at all wavelengths longer than 1 µm making IRIS *the* work-horse diffraction limited instrument for TMT.

---

[*] larkin@astro.ucla.edu; phone 1 310 825-9790; fax 1 310 206-2096

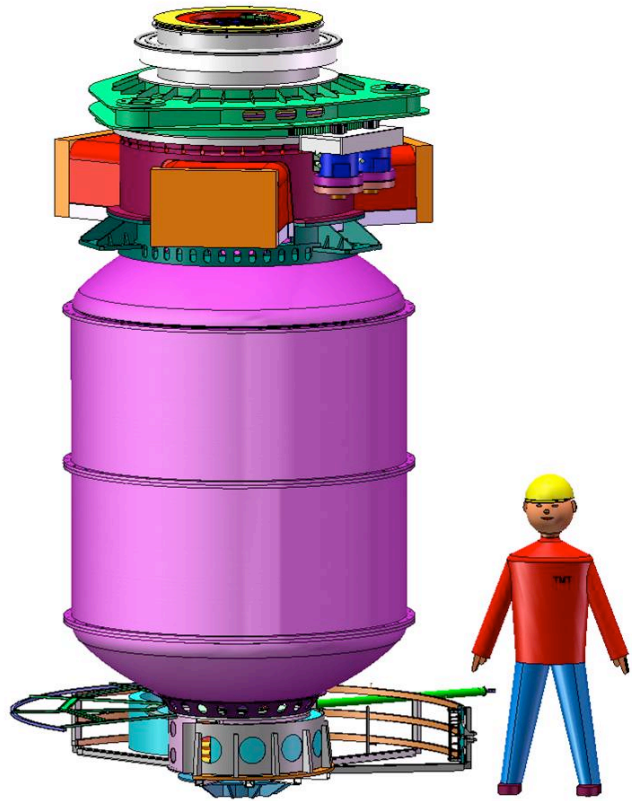

Figure 1: The entire IRIS instrument shown next to a 6 foot tall model human. Light from the AO system NFIRAOS enters from the top through the insulated snout. The entire instrument is suspended from the AO system and rotates in the position shown about its central axis with fixed gravity vector.

The science fields of view are given in table 1 below. The slicer spectrograph always has 45x88 spatial elements in its rectangular field of view, and the lenslet array spectrograph has a normal mode of 112x128 spatial elements and a secondary field of view of 16x128 spatial elements to increase the number of available spectral elements. The imager mode can be used in parallel and independently with any of the spectrograph modes.

Table 1. The fields of view and sample sizes of the different science modes

| Mode | Plate scale | Field of View | Number of Spectral Elements per Spaxel |
|---|---|---|---|
| Direct Imager | 4 mas | 16.4 x 16.4 arcsec | 1 |
| Slicer Spectrograph | 50 mas | 4.4 x 2.25 arcsec | 2000 |
| 90x45 Spaxels | 25 mas | 2.2 x 1.125 arcsec | 2000 |
| Lenslet Spectrograph | 9 mas | 1.01 x 1.15 arcsec | 500 |
| 112x128 Spaxels | 4 mas | 0.45 x 0.51 arcsec | 500 |
| Lenslet Spectrograph | 9 mas | 0.144 x 1.15 arcsec | 4096 |
| 16x128 Spaxels | 4 mas | 0.064 x 0.51 arcsec | 4096 |

## 1.1 Scientific Objectives

In the new era of extremely large telescopes like TMT, science will evolve extremely rapidly so it is difficult to predict what will be the most exciting scientific problems that IRIS will help to solve. However, our science team has demonstrated that a diffraction limited integral field spectrograph at TMT will open entirely new opportunities in virtually every area of astrophysical science. Table 2 below gives the physical scale that IRIS will be able to resolve at various astrophysical distances. So IRIS will not only resolve surface features 10's of kilometers across on Titan, it will also map the most distant galaxies at the scale of an individual star forming region. With a powerful spectral resolution of 4000 (75 km/sec) available in all plate scales and wavelengths, kinematic substructure in those same galaxies, and chemical compositions of moon surface features are attainable. The resolution of 4000 also resolves the individual night sky lines of the Earth's atmosphere allowing us to primarily work in the much darker continuum regions of the spectrum.

Table 2. The table below shows what a typical PSF core diameter maps to as a function of distance. The angular scales available to TMT with an AO system are truly tantalizing allowing us unprecedented resolution into solar system objects, nearby extrasolar systems, stellar clusters, distance galaxies and the early universe.

| Distance | Spatial Scale for 0.01" (Astrophysical Example) |
|---|---|
| 5 A.U. | = 36 km (Jovian planets and moons) |
| 5 pc | = 0.05 AU (Nearby stars - search for companions) |
| 100 pc | = 1 AU (Nearest star forming regions) |
| 1 kpc | = 10 AU (Typical Galactic objects) |
| 8.5 kpc | = 85 AU (Galactic Center or Bulge) |
| 1 Mpc | = 0.05 pc (Nearest galaxies) |
| 20 Mpc | = 1 pc (Virgo Cluster) |
| Z=0.5 | = 0.07 kpc (galaxies at the time of the sun's formation) |
| Z=1.0 | = 0.09 kpc (formation of disks, drop in global SFR) |
| Z=2.5 | = 0.09 kpc (QSO epoch, H$\alpha$ in K band) |
| Z=5.0 | = 0.07 kpc (protogalaxies, QSO's, epoch of reionization) |

Details of individual science cases are discussed in Barton et al.[7] and sensitivity and performance estimates are given in Wright et al.[8].

## 1.2 TMT and its Adaptive Optics System NFIRAOS

One of the most exciting aspects of the Thirty Meter Telescope will be its ability to achieve an unprecedented angular resolution for a single aperture. The TMT team has always seen adaptive optics and diffraction limited imaging as a core capability that should be available at first light. To make this a reality, an advanced multiconjugate adaptive optics system dubbed NFIRAOS[2] (Narrow Field Infrared Adaptive Optics System) is being designed in parallel with the telescope to be ready at TMT first light. NFIRAOS delivers a well-corrected two arcminute field of view and can take advantage of both natural and laser guide stars. Since there are two deformable mirrors, plus the ability to focus the telescope, the system can vary its effective focal length without affecting the image quality significantly at the science field. This has the unfortunate effect that without an independent set of reference points, the plate scale can vary in real time and lead to distortion and image smearing. To independently monitor and maintain the plate scale, and thus the image distortion, IRIS must have three tip/tilt wavefront sensors to track stars across the field of view. This is the role of the on-instrument wavefront sensors described below and in Loop et al.[3]. Rotating the full field of NFIRAOS with an optical rotator such as a K-mirror is also not feasible. So the science field of view is rotating at the NFIRAOS output ports. To correct for field rotation, the entire IRIS instrument is rotated in real-time and in closed loop with the AO system again using the wavefront sensors to maintain proper angular position.

## 2. OPTICAL SUBSYSTEMS

IRIS has three distinct optical units. In the front of the instrument are the on-instrument wavefront sensors (OIWFS) that can monitor up to three stars across a 2 arcmin region to remove image distortion and provide tip/tilt and focus correction. The science portion contains an on-axis "wide field" imager with a 16.4 x 16.4 arcsec field of view and a slightly off-axis integral field spectrograph with four separate plate scales. Both the imager and spectrograph can be used simultaneously and have independent filter wheels and optical components allowing for completely parallel observations. The integral field spectrograph is particularly ambitious spanning plate scales from 4 mas (Nyquist sampling at 1 micron) to a coarsest platescale of 50 mas designed to maximize sensitivity on low surface brightness sources like galaxies in the very early universe. To achieve this large range of scales, two different techniques are used. The two finest scales (4 and 9 mas) use a lenslet-based optical design to maximize field of view (112x128 spatial samples) and achieve the highest possible image quality. The coarsest two scales (25 and 50 mas) use a slicer-based design to provide more flexible spectral formats and greater spectral bandwidth (spectra up to 2000 pixels in length). The four IFU plate scales are concentric with each other and simple two position stages are used to divert the light into the separate optical paths with high positional accuracy. A coincidence of the slicer and lenslet design parameters makes an F/4 camera optimal for formatting the light onto the science detector. So before the light is split into either the slicer or lenslet paths, it passes through a common filter wheel, rotating pupil stop and ADC. Then light is recombined at the grating turret with a common set of gratings, camera optics and detector. This sharing of most of the optical elements leads to major cost savings in the instrument and actually would not be possible if a single slicer or lenslet had been attempted for all four plate scales.

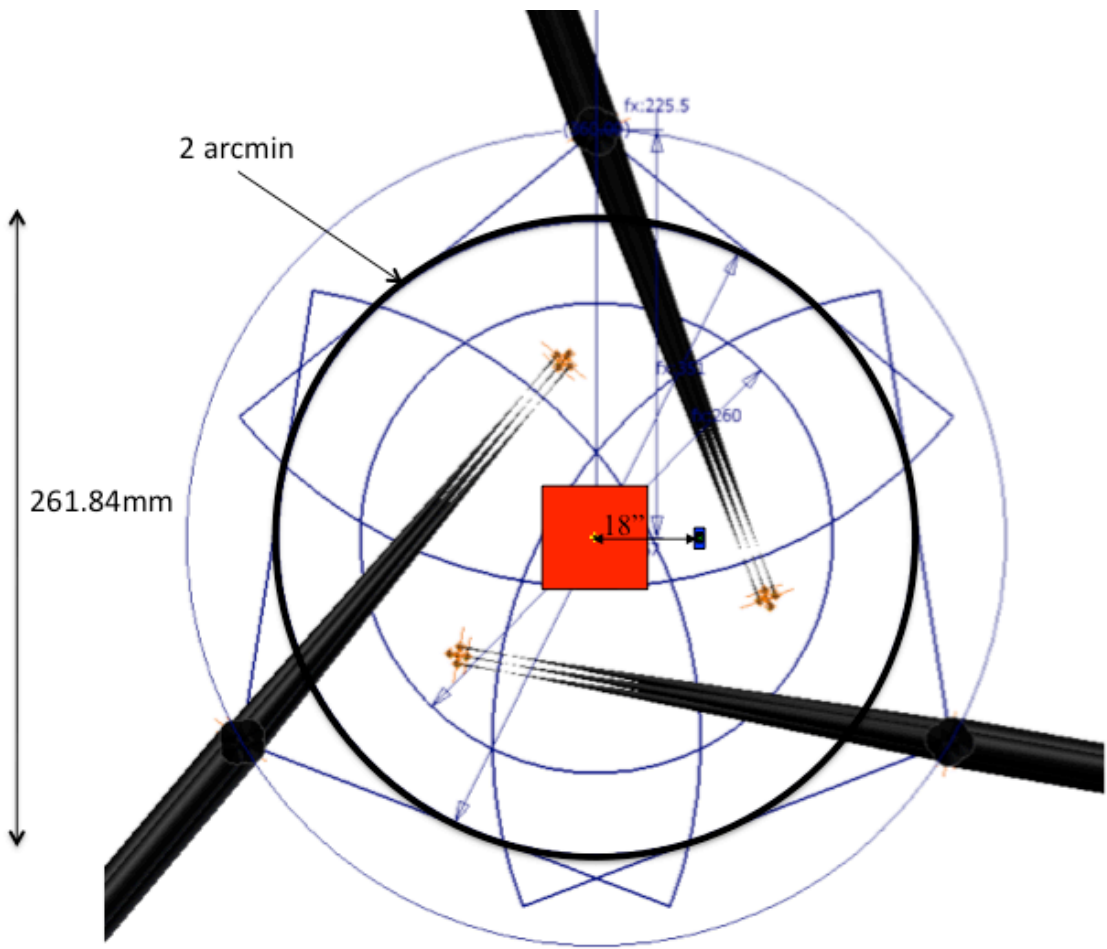

Figure 2: The division of the IRIS entrance field relative to the On-Instrument Wavefront Sensor probe arms. Entrance field is 2 arcmin in diameter and is patrolled by the 3 OIWFS probe arms. The red square is the on-axis, 16.4 x 16.4 arcsec imager FOV. The concentric lenslet and image slicer IFU fields (four channels in total) are offset by 18 arcsec compared to the NFIRAOS optical axis and are shown as small blue and green rectangles.

With the combination of lenslet and slicer spectrographs and a variety of gratings and filters, IRIS is capable of an extensive number of spectral products. The table below lists the currently planned grating suite along with its complimentary set of filters and field modes.

Table 3. The table below lists the currently planned gratings and operational modes of the instrument. In each case, at least two plate scales are supported. Any mode listed as "Slicer" is available for the 25 and 50 mas plate scales and has 45x88 independent field points. Those listed as "Lenslet 112x128" have 4 and 9 mas plate scales available and 112x128 field points. "Lenslet 16x128" also has 4 and 90 mas platescales, but the more limited 16x128 field points.

| # | Grating groove Density (per mm) | Blaze angle | Filters | Field mode | Resolution | Bandwidth |
|---|---|---|---|---|---|---|
| 1 | 341 | 9.86 | Zbb | Slicer | 4000 | 20% |
|   |     |      | Zn1-Zn4 | Lenslet 112x128 | 4000 | 5% |
| 2 | 290 | 9.86 | Ybb | Slicer | 4000 | 20% |
|   |     |      | Yn1-Yn4 | Lenslet 112x128 | 4000 | 5% |
| 3 | 249 | 9.86 | Jbb | Slicer | 4000 | 20% |
|   |     |      | Jn1-Jn4 | Lenslet 112x128 | 4000 | 5% |
| 4 | 194 | 9.86 | Hbb | Slicer | 4000 | 20% |
|   |     |      | Hn1-Hn5 | Lenslet 112x128 | 4000 | 5% |
| 5 | 145 | 9.86 | Kbb | Slicer | 4000 | 20% |
|   |     |      | Kn1-Kn5 | Lenslet 112x128 | 4000 | 5% |
| 6 | 161 | 9.86 | H+K | Lenslet 16x128 | 4000 | 2x20% |
| 7 | 709 | 20.54 | Zn1-Zn4 | Slicer | 8000 | 5% |
|   |     |       | Zbb | Lenslet 16x128 | 8000 | 20% |
| 8 | 606 | 20.54 | Yn1-Yn4 | Slicer | 8000 | 5% |
|   |     |       | Ybb | Lenslet 16x128 | 8000 | 20% |
| 9 | 516 | 20.54 | Jn1-Jn4 | Slicer | 8000 | 5% |
|   |     |       | Jbb | Lenslet 16x128 | 8000 | 20% |
| 10 | 398 | 20.54 | Hn1-Hn5 | Slicer | 8000 | 5% |
|    |     |       | Hbb | Lenslet 16x128 | 8000 | 20% |
| 11 | 310 | 20.54 | Kn1-Kn3 | Slicer | 8000 | 5% |
| 12 | 282 | 20.54 | Kn4-Kn5 | Slicer | 8000 | 5% |
| 13 | 296 | 20.54 | Kbb | Lenslet 16x128 | 8000 | 20% |
| 14 | 872 | 26.0 | Zbb | Lenslet 16x128 | 10,000 | 20% |
| 15 | 741 | 26.0 | Ybb | Lenslet 16x128 | 10,000 | 20% |
| 16 | 637 | 26.0 | Jbb | Lenslet 16x128 | 10,000 | 20% |
| 17 | 497 | 26.0 | Hbb | Lenslet 16x128 | 10,000 | 20% |
| 18 | 368 | 26.0 | Kbb | Lenslet 16x128 | 10,000 | 20% |
| 19 | TBD | TBD | TBD | Lenslet 16x128 | Up to 40,000 | 5% |
| 20 | Mirror | 22.5 | Any | Any | NA | NA |

**2.1 Lenslet Array Integral Field Spectrograph**

Among the greatest challenges of the instrument is the preservation of the excellent image quality provided by the AO system in the finest plate scales. To ensure no more than a 10% loss in Strehl ratio, the non-correctable wavefront error to the science focal plane must be less than 30 nm rms. In an integral field spectrograph, this is particularly challenging since the AO system cannot normally correct for light that follows different paths through the spectrograph. As demonstrated by OSIRIS[9], a spectrograph based on a lenslet array can preserve image quality at the level of 25 nm of rms wavefront error by sampling the image plane prior to the spectrograph optics. So in IRIS we elected to use a lenslet based design for the two finest plate scales (4 and 9 mas per sampling). The only non-common path optics that affect image quality are then the input window, two air-spaced doublets, a filter and a fold mirror, all of which can be partially corrected through image sharpening with the AO system.

The lenslet array has a square pattern and a pitch of 350 microns per lenslet. The array is placed in the focal plane of simple reimaging optics and the pupil produced by each lenslet is typically 32 times smaller than its own diameter including the effects of pupil diffraction. This compression is sufficient to allow 16 spectra to interleave together in the projected height of an individual lenslet. So a basic pattern of 16x16 lenslets is used as a building block for the full field.

A 7x8 pattern of these basic blocks are then used to create the 112x128 spatial samples. Each lenslet's spectrum can extend for the projected length of a 16x16 block or approximately 512 pixels. In the dominant modes where the spectral resolution is 4000, this corresponds to a 5% bandpass in each individual exposure. Figure 3 shows an example of how the spectra fall on the detector. Each horizontal stripe is the spectrum of one lenslet; in this case a 5% bandpass spectrum near 0.95 microns. The bands of stripes running diagonally down and to the right are each from a row of lenslets. So every row produces one of these bands. So you can see the individual spectra, only the lower 500 pixel rows of the detector are shown. For more discussion of the Lenslet optical design see Moore et al.[5].

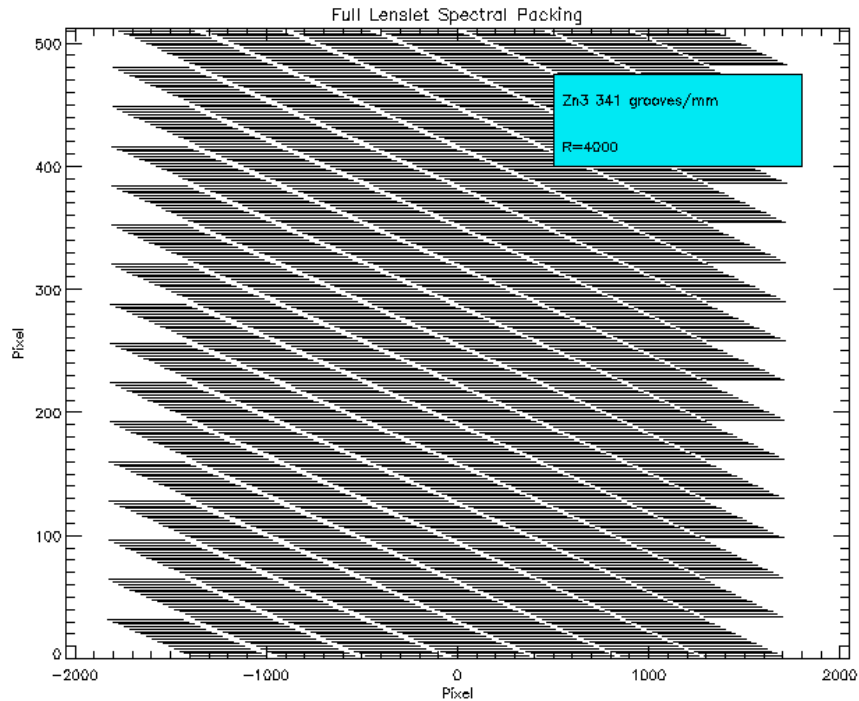

Figure 3: This figure above shows how the 14,336 spectra are typically arranged on the 4096x4096 pixel detector (only the bottom 500x4096 pixels are shown and stretched to see the individual spectra). In this example, a 341 grooves/mm grating is used with a Z-band narrow filter (Zn3:0.91-0.96 microns). The full 112x128 lenslet array is illuminated. The spectral resolution is 4000 which is the most common mode planned.

**2.2 Slicer Integral Field Spectrograph**

A mirror slicer has the advantage over a lenslet based spectrograph of reformatting the focal plane into approximately linear feeds to the spectrograph portion with fairly arbitrary spacing. In the case of IRIS, we will use 88 flat mirror slicers arranged in a stack to initially cut up the focal plane. The resulting 88 beams are then reorganized into two linear feeds to the spectrograph portion. If we think of these slit-like feeds as being assigned pixel real estate on the quadrants of the detector with spectral dispersion running horizontally, then one strip subtends the two left quadrants and the other the two right quadrants. So spectra can have up to 2000 spectral pixels. Figure 4, shows a schematic generated from an IDL model of the instrument with the approximate use of the detector by the slicer spectra. In this figure each horizontal stripe is the spectrum from one spaxel. All of the spaxels from 44 of the slicers occupy the left set of spectra, and the others are on the right. A benefit of the slicer design is that adjacent spectra have nearly identical wavelengths next to each other making data reduction easier and cross contamination less of a problem.

For IRIS, the slicer consists of 88 mirror facets each projected to 90 pixels in length (45 resolution elements) and 2 pixels in width. Pairs of spherical mirrors are used after the initial stack of flat slicers to perform the reorganization of the focal plane into the linear feeds. The spherical mirrors also server to de-magnify the slicer facets so the resulting linear inputs are six times shorter in physical length than the original facets aligned end-to-end. A refractive set of lenses after the slicer output focal plane is used to collimate the light onto the gratings. For details of the slicer design see Moore et al.[5].

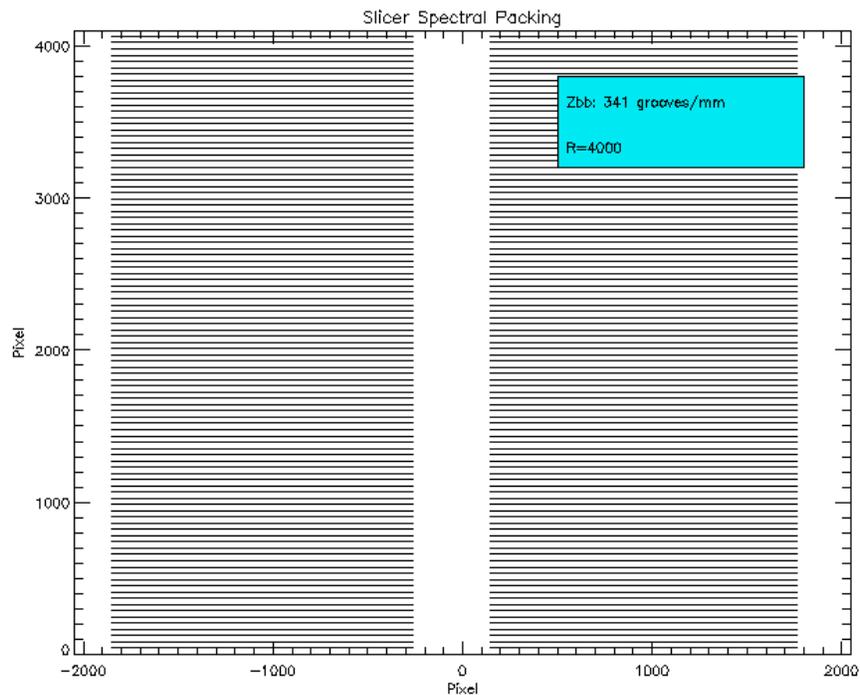

Figure 4: Like figure 3, the figure above shows how the slicer spectra are rearranged on the 4096x4096 pixel detector. The same grating (341 grooves/mm) is used, but in this case a broad band Z filter is used. The spectral resolution is again 4000 which is the most common mode planned. Unlike the lenslet spectra, adjacent spatial locations have their spectra aligned so the same wavelengths are adjacent.

**2.3 Imager**

The imager is designed to achieve the sample diffraction limit of TMT at all IRIS wavelengths beyond 1 micron with a goal of 30 nm of rms wavefront error. With the expectation of detectors with 4096 pixels on a side (16 megapixels), and sampling with 4 milliarcseconds per pixel, we have designed optics to provide this image quality over a 16.4x16.4 arcsecond field. We looked at many configurations using relective and refractive elements and found that a refractive option based on apochromatic triplets best met the requirements. The optical system consists of a collimator and camera both with a $BaF_2$-Fused Silica-ZnSe apochromatic triplet and a single $BaF_2$ lens near the focus. The rms wavefront error of the system is less than 22 nm with ideal optical parameters. A sensitivity analysis shows that a reasonable amount of errors in fabrication and alignment will give the rms wavefront error of less than 30 nm in 90 % of all cases. We find that given expected mechanism repeatability, distortion errors are well below 100 microarcseconds. For more information on the imager see Suzuki et al.[4].

**2.4 On Instrument Wavefront Sensors (OIWFS)**

The NFIRAOS adaptive optics system will often use laser guide stars and perform a multi-conjugate analysis of the atmosphere. This allows it to correct stellar images over a full 2 arcminute field, but the LGS AO system itself is blind to several of its own corrective modes. In particular, field distortion, tip/tilt and focus must be determined independently using real stars and natural guide star sensors. For full correction of these effects, three stars must be monitored in the corrected field of view. Given the density of stars on the sky, achieving full correction over at least half of the sky (or equivalently for half of anyone's targets of interest) requires using very faint stars where AO correction is needed to improve their contrast against the background light. As a result, IRIS is designed with three pick-off arms which patrol the AO corrected field of view of NFIRAOS and which will use infrared wavefront sensors to gain from the high Strehl ratios in the infrared. To further improve the sensitivity, the sensors will integrate over a very wide wavelength range (0.84-1.7 microns). The sensors are cryogenic infrared detectors probably cooled to liquid nitrogen temperatures, but the optical components including probe arms must be cooled to roughly -30 C to reduce the thermal background.

Since the OIWFS serves as the direct measurement of stars on the sky, its performance is the dominant factor in the astrometric performance of the system. In other words, if one of the probes is misplaced or makes a poor centroid measurement, then the wavefront sensors will command the AO system to move that star to the wrong position, which will both distort and blur the image delivered to the science cameras. Since there is differential atmospheric dispersion between the LGS wavefront sensors wavelength range, the OIWFS wavelength range, and the science wavelength, the stellar images are also moving differentially in real time from the optical position of the stars, and with respect to the science target. As a result, the probe arms must move very accurately (typical rms of better than 4 microns) and must track well as stars transition across pixel boundaries.

## 2.5 Atmospheric Dispersion Compensation

The angular resolution of IRIS turns some traditionally minor inconveniences into major problems. Atmospheric dispersion is a particularly problematic concern at TMT's spatial resolution and is something the team has worked hard to understand; an analysis is presented in detail in Phillips et al.[10]. As illustrated in figure 5 below, a star's position, even across a 20% bandpass filter, can vary by many resolution elements from the blue end of the bandpass to the red end. In the figure, the difference in angular position between the bluest wavelength and reddest wavelength in each band is plotted against zenith distance to the star down to the zenith limit of 65 degrees. At 1 micron (Y-band), the diffraction limited core has a fwhm of roughly 0.008 arcseconds, but at zenith angles around 20 degrees, the differential dispersion is already more than 0.03 arcseconds or roughly 4 resolution elements. The direction of the dispersion is also always oriented along the elevation axis, so it sweeps out an arc over time compared to RA and dec. So even in an integral field spectrograph, where independent images are maintained at fine wavelength steps, a star's position cannot be held fixed in more than one wavelength slice at a time without dispersion compensation. In order to perform astrometry at the tens-of-microarcseconds level, real time and highly accurate dispersion correction is needed. To achieve this, we have looked at hundreds of glass combinations to find ones that work well over the full wavelength range, both for the OIWFS probes and for the science instruments. For reasons discussed in Phillips et al.[10], we are using crossed Amici prisms to correct for the atmospheric dispersion. The counter-rotating prism pairs can produce a stellar image with less than 1 milliarcsecond of residual motion across any of our bandpasses, and with knowledge at a significantly higher level. We've also investigated dispersion issues of field distortion introduced by the atmosphere, centroid shifts that can occur due to the color of the tip/tilt stars, and the effect of uncertain atmospheric conditions (pressure and temperature) at the telescope at the time of an exposure.

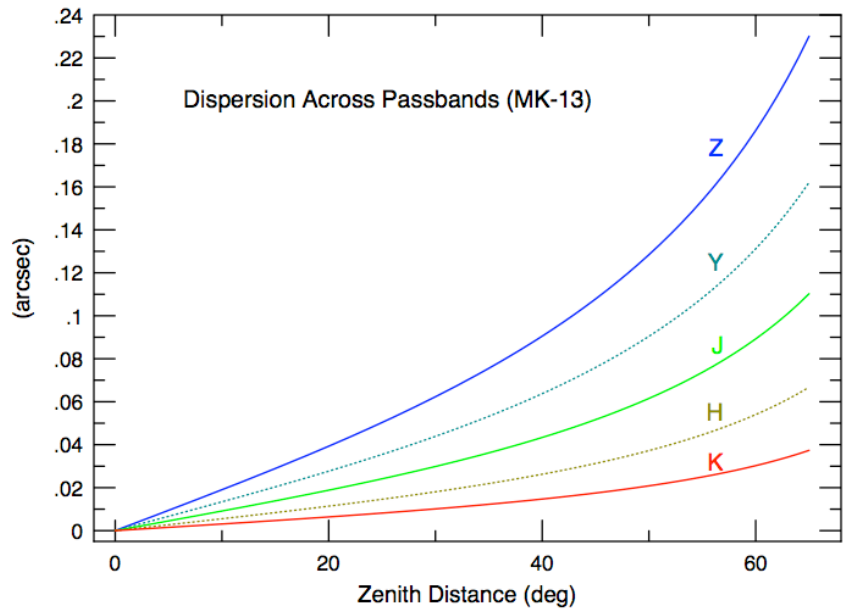

Figure 5. Atmospheric dispersion within the various passbands covered by IRIS, as a function of zenith distance. This figure shows the dispersion of the blue edge relative to the red edge of each passband. The dispersion has been calculated for the adopted site of TMT on Mauna Kea.

# 3. MECHANICAL DESIGN

IRIS is a very large cryogenic instrument that must work reliably night after night. Thermal and mechanical stability are also crucial for the performance of the instrument and dramatically affect the ability to reduce and analyze the complex data products. In all aspects of the design, we have started with existing reliable solutions, especially in the area of cryogenic mechanisms and optical mounts.

## 3.1 Vacuum Cryogenic System

The vacuum vessel is approximately 2 meters in diameter and 3.5 meters long. This is comparable to the diameter of the MOSFIRE instrument[11] (PI: McLean), which is a collaborative project between UCLA, Caltech, UCSC and the Keck Observatory. As with MOSFIRE, we plan on concave end caps for the dewar and cylindrical sections that allow us to assemble it sequentially. The figure below shows the vacuum chamber in comparison to the MOSFIRE instrument.

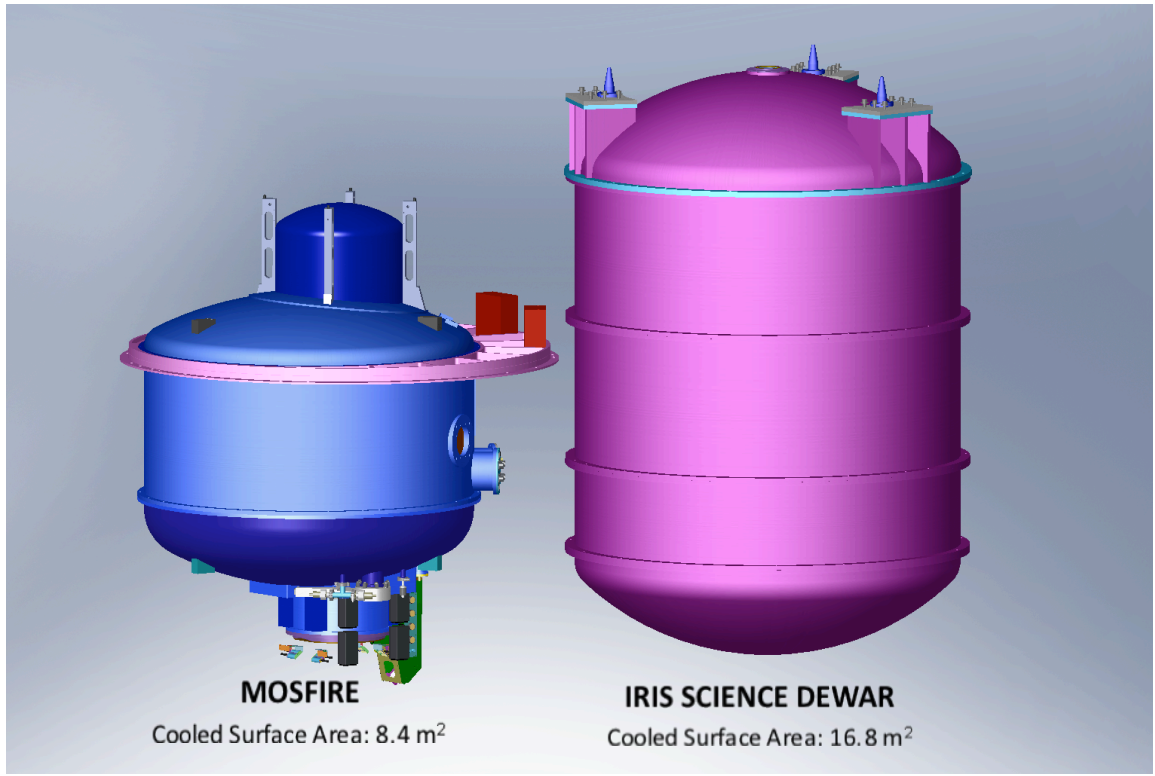

Figure 6. The MOSFIRE instrument is close to delivery at the Keck telescope and its dewar serves as a good comparison for the proposed IRIS vacuum chamber. Domed end caps and several separate sections provide access to the internal components.

Inside the vacuum chamber, the optical benches must provide extremely stable optical platforms, while remaining thermally isolated from the chamber itself. As shown in the cut-away below, the heavily light-weighted optical benches are joined by a thin shell to form a rigid structure. It, in turn, is attached at several points by very low conduction G10 fiberglass struts. Unlike MOSFIRE, IRIS works in a fixed gravity orientation, so flexure of this large structure is not a major concern. Access during assembly and operation, however, requires a great deal of forethought and planning and a series of hatches and access points are planned.

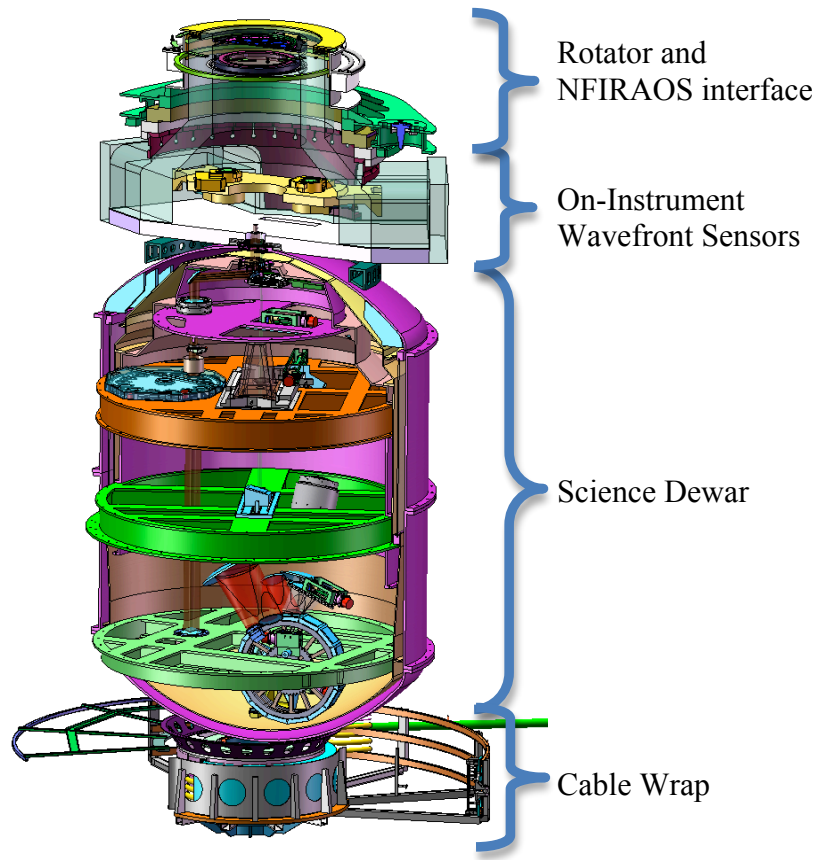

Figure 7. This cut-away of the entire instrument shows the major optical benches, optical elements and mechanisms. A thin cold shield provides a rigid internal structure to support the benches and to attach through fiberglass to the outer vacuum chamber.

### 3.2 Field Rotation

The instrument mounts to the lower port of the NFIRAOS instrument and can rotate in place without changing the gravity orientation of the instrument. However, rotating such a large cryogenic instrument with microarcsecond astrometric precision is very challenging. Shown above in figure 7, the instrument rotator is at the top of the dewar and is integrated with the wavefront sensor package. The large rotary bearing provides support for the instrument and is also thermally insulated from the NFIRAOS. A snout extends into the NFIRAOS output port to provide an insulated connection between the -30 C NFIRAOS chamber and the -30 C OIWFS section. In the current design, a cable wrap at the bottom of the instrument allows all of the umbilical connections to wrap around the instrument as it rotates without hindering the field rotation.

### 3.3 Cryogenic Mechanisms

As with any cryogenic instrument, especially at the size scale of IRIS, mechanisms should be minimized and must be extremely robust (typically with 10-year demonstrated life). For the science dewar, where mechanisms will likely operate between 77 and 120 Kelvin, the following mechanisms are currently planned:

Imager Mechanisms (6 axes of control):

- Atmospheric dispersion compensation system – two sets of counter-rotating prisms – continuous motion
- Rotating pupil mask – simple rotation but with continuous motion

- Filter wheel with roughly 36 positions
- Pupil viewing selection mirror – two position stage
- Detector focus – micron level motion control

Spectrometer Mechanisms (10 axes of control):

- Atmospheric dispersion compensation system – two sets of counter-rotating prisms – continuous motion
- Rotating pupil mask – simple rotation but with continuous motion
- Filter wheel with roughly 36 positions
- Slicer or lenslet selector – two position stage with fold mirror
- Two plate scale stages – in each spectrometer type, there is a simple 2-position stage to select plate scale
- Lenslet version will require a slit scanning mechanism for calibration
- Grating turret with at least 10 positions – requires micron level repetition
- Detector focus – micron level motion control

In many ways, the delivered mechanisms in NIRSPEC, OSIRIS, MOSFIRE, and GPI represent baseline designs for the mechanisms above. For each of the 16 cryogenic mechanisms we have started with an existing cryogenic stage or wheel and modified it for the IRIS application. Only in a few cases, such as the grating turret, is a particular specification (in this case angular repeatability) more stringent than in previous mechanisms. In such cases, we plan on prototyping during the preliminary design phase.

## 4. DETECTOR

The design of the instrument is contingent on the availability of new infrared detector arrays with 16 Megapixels (4096x4096 format) from Teledyne[6]. Our expectation is to have devices with read noise of order 2-3 electrons after multiple samples and a dark current well below 0.01 electrons per second. Since Teledyne is developing two versions of 4Kx4K devices with 10 and 15 micron pitch pixel spacing, respectively, we have developed imager designs around both detector sizes. The spectrographs would struggle with the smaller pitch, so it is baselined to use the 15 micron devices.

As with the current Hawaii-2RG from Teledyne, we plan to use the Teledyne ASIC processor to control the detector and process analog signals. This greatly simplifies the electronics requirements of the instrument and provides a cryogenic set of preamplifiers and digitization greatly minimizing noise.

## 5. SUMMARY

We have developed a conceptual design for an integral field spectrograph and imager that will work at the diffraction limit of the Thirty Meter Telescope. We have working designs for all components and subsystems of the spectrograph and are preparing to enter a preliminary design phase. The instrument builds on the heritage of several previous instruments including the Keck instruments OSIRIS[9] and MOSFIRE[11].

### 5.1 Future work

In the development of the conceptual design for IRIS we have identified several challenging subsystems and components. Many of the difficulties arise from the need to satisfy stringent requirements at cryogenic temperatures. Among the items we'd like to prototype or explore in the next phase are:

1. Prototype probe arms for the On-Instrument Wavefront sensors (OIWFS): The position and knowledge of the position of these arms serves as "truth" to the adaptive optics system so their performance directly impacts distortion, astrometry and image core size.

2. Grating turret: Unlike OSIRIS, we are proposing to use multiple gratings in a turret. In any integral field spectrograph data reduction is a major concern, and spectral repeatability is crucial. So our goals include roughly micron level repeatability of this large cryogenic mechanism in both axes.

3. Three mirror anastigmats: To meet the field of view and image quality requirements of the spectrograph, we believe that high order aspheric mirrors made from a low expansion glass such as zerodur are required. But aligning such a system to work cryogenically in a metal housing will require a new level of deterministic assembly with coordinate measuring machines.
4. Spinel glass: One of the most promising glasses for atmospheric dispersion correction is the hot pressed material known as Spinel. Primarily produced in large sizes as a bulletproof glass, its optical properties are still somewhat uncertain, especially its scattering properties. We plan on obtaining and testing blanks of this material.

### 5.2 Schedule

The IRIS project is currently in a conceptual design phase, which was preceded by a feasibility study that ended in 2008. We are planning the conceptual design review in October 2010 and hope to begin a preliminary design study in January 2011. The following list gives our anticipated future milestones leading up to first light with TMT and NFIRAOS.

- Oct. 2010              Conceptual Design Review
- Jan 2011-Dec 2011      Preliminary Design Phase
- Jan 2012-June 2013     Final Design Phase
- July 2013-July 2015    Fabrication, Assembly and Subsystem Testing
- July 2015-Apr 2016     Science Dewar Integration – OIWFS goes to work with NFIRAOS
- Apr 2016-Oct 2016      Full Performance Testing of Science Dewar
- Dec 2016-Aug 2017      Integrate Science Dewar, OIWFS and NFIRAOS at HIA
- Oct 2017-Aug 2018      Integrate with Observatory
- Aug 2018               Begin commissioning with TMT
- Nov 2018               FIRST LIGHT for TMT with NFIRAOS and IRIS

## ACKNOWLEDGMENTS

The authors wish to acknowledge the support of the TMT partner institutions. They are the Association of Canadian Universities for Research in Astronomy (ACURA), the California Institute of Technology and the University of California. This work was supported as well by the Gordon and Betty Moore Foundation, the Canada Foundation for Innovation, the Ontario Ministry of Research and Innovation, the National Research Council of Canada, Natural Sciences and Engineering Research Council of Canada, the British Columbia Knowledge Development Fund, National Astronomical Observatory of Japan (NAOJ), the Association of Universities for Research in Astronomy (AURA) and the U.S. National Science Foundation.